\documentclass[12pt,epsf]{article}
\usepackage{verbatim}

\usepackage{dsfont}
\usepackage{sidecap}
\sidecaptionvpos{figure}{t}
\usepackage{subfigure}
\usepackage{enumerate} 

\usepackage{amssymb,amsmath}
\usepackage{graphicx, xcolor, varwidth}
\usepackage{setspace}
\usepackage[permil]{overpic}
\usepackage{cite}

\usepackage{feynmp}
\colorlet{darkblue}{blue!70!black}
\colorlet{darkgreen}{green!70!black}

\usepackage[colorlinks=true,urlcolor=darkblue,linktocpage=true,linkcolor=darkblue,citecolor=darkblue]{hyperref}

\numberwithin{equation}{section}

\newcommand{\be}{\begin{equation}}
\newcommand{\ee}{\end{equation}}
\newcommand{\bea}{\begin{eqnarray}}
\newcommand{\eea}{\end{eqnarray}}
\newcommand{\bear}{\begin{eqnarray}}
\newcommand{\eear}{\end{eqnarray}}  
\newcommand{\beas}{\begin{eqnarray*}}
\newcommand{\p}{\partial}
\newcommand{\eeas}{\end{eqnarray*}}
\newcommand{\ba}{\begin{array}}
\newcommand{\ea}{\end{array}}

\newcommand{\del}{\nabla}

\newcommand{\pd}[2][1]{\ifnum#1=1 \frac{\partial}{\partial {#2}} \else
  \frac{\partial^#1}{\partial {#2}^{#1}}\fi}
\newcommand{\dpd}[2][1]{\ifnum#1=1 \dfrac{\partial}{\partial {#2}} \else
  \frac{\partial^#1}{\partial {#2}^{#1}}\fi}
\newcommand{\td}[2][1]{\ifnum#1=1 \frac{d}{d{#2}} \else
  \frac{d^#1}{d{#2}^{#1}}\fi}

\newcommand{\nbox}{{\,\lower0.9pt\vbox{\hrule \hbox{\vrule height 0.2 cm \hskip 0.19 cm \vrule height 0.2 cm}\hrule}\,}}

\newcommand{\ie}{{\it i.e.,}\ }

\newcommand{\btau}{\bar{\tau}}

\newcommand{\half}{\tfrac{1}{2}}
\newcommand{\bm}{\bar{m}}
\newcommand{\bz}{\bar{z}}



\newdimen\tableauside\tableauside=1.0ex
\newdimen\tableaurule\tableaurule=0.4pt
\newdimen\tableaustep
\def\phantomhrule#1{\hbox{\vbox to0pt{\hrule height\tableaurule width#1\vss}}}
\def\phantomvrule#1{\vbox{\hbox to0pt{\vrule width\tableaurule height#1\hss}}}
\def\sqr{\vbox{%
  \phantomhrule\tableaustep
  \hbox{\phantomvrule\tableaustep\kern\tableaustep\phantomvrule\tableaustep}%
  \hbox{\vbox{\phantomhrule\tableauside}\kern-\tableaurule}}}
\def\squares#1{\hbox{\count0=#1\noindent\loop\sqr
  \advance\count0 by-1 \ifnum\count0>0\repeat}}
\def\tableau#1{\vcenter{\offinterlineskip
  \tableaustep=\tableauside\advance\tableaustep by-\tableaurule
  \kern\normallineskip\hbox
    {\kern\normallineskip\vbox
      {\gettableau#1 0 }%
     \kern\normallineskip\kern\tableaurule}%
  \kern\normallineskip\kern\tableaurule}}
\def\gettableau#1 {\ifnum#1=0\let\next=\null\else
  \squares{#1}\let\next=\gettableau\fi\next}

\newcommand{\smallstart}{
\end{spacing}
\noindent\hfil\rule{1\textwidth}{.4pt}\hfil\small

   \addtolength{\leftskip}{5mm}
}
\newcommand{\smallend}{
   \addtolength{\leftskip}{-5mm}
\noindent\hfil\rule{1\textwidth}{.4pt}\hfil\normalsize
\begin{spacing}{1.3}
}

\begin{document}
\begin{spacing}{1.3}
\begin{titlepage}

\begin{center}
{\Large \bf
Fast Conformal Bootstrap  \\ \vspace {0.5cm}
 and Constraints on 3d Gravity
}

\vspace*{12mm}

Nima Afkhami-Jeddi, Thomas Hartman, and Amirhossein Tajdini

\vspace*{6mm}

\textit{Department of Physics, Cornell University, Ithaca, New York\\}

\vspace{6mm}

{\tt na382@cornell.edu, hartman@cornell.edu, at734@cornell.edu}

\vspace*{15mm}
\end{center}
\begin{abstract}

The crossing equations of a conformal field theory can be systematically truncated to a finite, closed system of polynomial equations.  In certain cases, solutions of the truncated equations place strict bounds on the space of all unitary CFTs.  We describe the conditions under which this holds, and use the results to develop a fast algorithm for modular bootstrap in 2d CFT.  We then apply it to compute spectral gaps to very high precision, find scaling dimensions for over a thousand operators, and extend the numerical bootstrap to the regime of large central charge, relevant to holography. This leads to new bounds on the spectrum of black holes in three-dimensional gravity. We provide numerical evidence that the asymptotic bound on the spectral gap from spinless modular bootstrap, at large central charge $c$,  is $\Delta_1 \lesssim c/9.1$.

\end{abstract}

\end{titlepage}
\end{spacing}

\vskip 1cm

\setcounter{tocdepth}{1}
\tableofcontents
\addtocounter{page}{1}

\begin{spacing}{1.3}\section{Introduction}

The conformal bootstrap is a rapidly growing array of techniques to solve strongly interacting quantum field theories. It had early success in two spacetime dimensions \cite{Belavin:1984vu}, and more recently has proved successful in higher dimensions \cite{Poland:2018epd}. An important impetus has been the development of numerical techniques to analyze the crossing equations, starting with the introduction of the functional bootstrap method in 2008 by Rattazzi, Rychkov, Tonni, and Vichi \cite{Rattazzi:2008pe}. This method uses linear (or semidefinite \cite{Poland:2011ey}) programming to constrain solutions to the crossing equations, thereby carving out the allowed parameter space of unitary CFTs. It has also been used to solve individual CFTs, including the 3d critical Ising model \cite{ElShowk:2012ht,El-Showk:2014dwa}. The method relies on finding extremal functionals, often numerically, which bound the space of CFTs, and in certain cases encode the spectrum of the target theory \cite{Poland:2010wg,ElShowk:2012hu}.

Despite this exciting progress, the current numerical methods are in their infancy, and there are many potential areas for improvement, both technical and conceptual. For example, with semidefinite programming, there is no clear way to leverage analytic results, such as the spectrum at high spin, to improve the numerics \cite{Simmons-Duffin:2016wlq}.  This is a tantalizing hint that much more efficient algorithms are waiting to be discovered. Such algorithms, combined with powerful analytics, will be needed to address certain physics problems, such as solving theories with large global symmetries \cite{Kos:2016ysd,Chester:2016wrc}, pushing bootstrap into the hydrodynamic regime \cite{Iliesiu:2018fao, Delacretaz:2018cfk}, or exploring the boundaries of quantum gravity by combining numerical bootstrap with AdS/CFT in the black hole regime. 

In this paper, we develop a new algorithm for numerical bootstrap, and apply it to  3d gravity.  The method is based on solving truncated crossing equations. It builds on one of the extremal functional methods of El-Showk and Paulos \cite{El-Showk:2016mxr}, and is very similar in many respects, but differs in some important details that extend the reach by several orders of magnitude. In principle, these methods are general, but in practice, they are limited to a certain class of bootstrap problems, where one can easily generate approximate solutions to the truncated equations. We will show that this is possible for the modular bootstrap problem relevant to 3d gravity.

The modular bootstrap is a variant of the conformal bootstrap introduced by Hellerman to constrain the spectrum of 2d CFTs \cite{Hellerman:2009bu}. Hellerman used it to prove analytically that any theory of 3d gravity has a primary operator with scaling dimension $\Delta \lesssim c/6$. From a holographic point of view, this bound is perhaps weaker than expected, because all known theories of 3d gravity have BTZ black holes \cite{Banados:1992wn} starting at $\Delta \sim c/12$. The bound can be systematically improved by numerics, but the difficulty grows rapidly with $c$ \cite{Friedan:2013cba,Collier:2016cls}. Previous numerical calculations were effective only for $c \lesssim 100$ and did not reach the semiclassical regime --- the bound had not converged in $c$, and it was not possible to estimate the asymptotics. A stronger bound is worth pursuing, because if it could be pushed down to the black hole threshold, then the numerics could also potentially be used to construct or rule out candidate theories of 3d gravity, or to suggest new analytic methods.

Using our algorithm, we extend the modular bootstrap up to $c \sim 1800$. This central charge appears to be in the semiclassical regime, where the bound has nearly converged, and we estimate the asymptotic bound to be $\Delta \lesssim c/9.1$. This estimate relies on an extrapolation in $c$, which appears to be reliable, but strictly speaking it is impossible to rule out a stronger bound as $c \to \infty$.  Our bound is not at the black hole threshold, so unfortunately cannot be used to construct candidate theories of 3d gravity or to address the questions above. We have not included the constraints from spin; it is possible that incorporating spin would lead to a stronger bound that saturates the black hole threshold.

The functional bootstrap is an expansion in derivatives, truncated at some order $N_{\rm deriv}$. Our calculations are performed up to derivative order $N_{\rm deriv} \sim 5000$. For a given $c$, this takes several CPU-hours on a standard laptop. It is impossible to make a direct comparison to existing methods, but a rough extrapolation suggests that semidefinite programming would require at least $10^7$ CPU-hours to perform the same calculation. A basic version of our algorithm, sufficient for modular bootstrap, is straightforward to implement in Mathematica with a few dozen lines of code. 

On the down side, our methods do not generalize in any obvious way to bootstrap problems with spin. This has (so far) prevented us from applying the algorithm to the spinning modular bootstrap, or to the 3d critical Ising model.

We begin with a review of the crossing equations and a summary of the truncation in section \ref{s:primal}. We explain under what circumstances the truncated equations can be used to place general bounds. In section \ref{s:dual} we review the functional method,  introduce a new method to optimize bootstrap functionals, and show that it is Lagrange dual to the truncation method. Section \ref{s:newmodular} describes the algorithm, and in section \ref{s:results} it is applied to modular bootstrap and 3d gravity.

Comparing our results to semidefinite programming requires very accurate bounds. In a standalone appendix \ref{app:secant}, we describe a method to find high precision bounds with linear programming, which has some advantages over the standard bisection method.

Although we have stressed the computational aspects, it seems likely that the truncated crossing equations can also be solving analytically in various limits --- lightcone \cite{Komargodski:2012ek,Fitzpatrick:2012yx}, large $N$ \cite{Heemskerk:2009pn}, short distance \cite{Mukhametzhanov:2018zja}, etc. This could be used to derive analytic bounds,  to jumpstart the numerics, or to develop variational methods that go to much higher orders. We  consider it likely that with more refinement, numerical methods can be extended to functionals with millions of derivatives. Further discussion of the outlook, including the challenges presented by problems with spin, is in section \ref{s:discussion}.

\section{Truncating the Primal Bootstrap}\label{s:primal}

The functional bootstrap is an optimization problem, and as such, it has a \textit{primal} and \textit{dual} formulation. The primal bootstrap refers to the crossing equations themselves. The dual approach uses linear functionals. We will rely on both pictures, starting on the primal side. In this section and section \ref{s:dual}, we consider a broad class of spinless bootstrap problems. We will specialize later to modular bootstrap.

\subsection{Setup}

Consider a crossing equation of the form
\be\label{cross1}
\sum_{\Delta,\ell} a_{\Delta}^\ell W_{\Delta}^\ell(z,\bz) = 0
\ee
where $\ell$ is spin, and $W_{\Delta, \ell}$ is the difference of a conformal block and a crossed block,
\be
W_{\Delta}^\ell (z,\bz) = g_{\Delta,\ell}(z,\bz) - g_{\Delta,\ell}(1-z,1-\bz) \ .
\ee 
Unitarity requires $a_{\Delta}^\ell  > 0$, and places lower bounds on $\Delta$.
Here we use the notation of the correlator bootstrap, with $z$ and $\bz$ conformal cross ratios, but the equations apply also to modular bootstrap, by replacing $(z,1-z) \to (\tau,-1/\tau)$ with $\tau$ the torus modulus. 

Following \cite{Rattazzi:2008pe}, let us act on \eqref{cross1} by linear functionals
\be
\mathcal{F} = \sum_{m=0}^M \sum_{\bm=0}^{\bar{M}} \alpha_{m,\bm} \left. \frac{\p^m}{\p z ^m} \frac{\p^{\bm} }{\p \bz^{\bm} } \right|_{z = \bz = 1/2}  \ .
\ee
(The methods may also apply to more general bases; see discussion section.) 
Choose a set $\{ \mathcal{F}_i \}_{i=1\dots P}$ of $P$ such functionals, and let
\be
f_i^\ell(\Delta) \equiv \mathcal{F}_i[ W_{\Delta}^\ell (z,\bz) ] \ ,
\ee
where for each $\ell$, the $f^\ell_i$ are linearly independent. Applying $\mathcal{F}_i$ to the crossing equation \eqref{cross1} gives 
\be\label{crossf}
\sum_{\Delta,\ell}a_{\Delta}^\ell f_i^\ell(\Delta) = 0 \ ,
\ee
(assuming the ${\cal F}_i$ are sufficiently well behaved to bring the derivatives inside the sum).
These are the \textit{primal} bootstrap equations. There are $P$ equations for an infinite number of parameters: the spectrum of $(\Delta,\ell)$ and the coefficients $a_{\Delta}^\ell $. After absorbing a positive prefactor into the $a_\Delta^{\ell}$, these equations are well approximated by polynomials \cite{Poland:2018epd}.

We will restrict our attention to bootstrap problems without spin dependence. In this case we set $z=\bz$ (or $\tau = -\btau$ in the modular bootstrap), and drop the spin label $\ell$. This class of problems includes spinless modular bootstrap, 1d correlators, the $sl(2)$ subsector of correlator bootstrap in higher dimensions.
\subsection{Truncation}

Let us take $P$ even, and truncate to $P/2 + 1$ states, including the identity term with $a_0 \equiv 1$, $\Delta_0 \equiv  0$. This results in the \textit{truncated primal equations}:
\be\label{primal}
\sum_{\mu = 0}^{P/2} a_{\mu} f_i(\Delta_\mu) = 0 \ .
\ee
This is now a closed system of $P$ polynomial equations for $P$ unknowns.

This seems a rather brutal truncation of the crossing equations, so at this point, it is not clear that it is useful. Nonetheless we will show that, sometimes, solutions of \eqref{primal} can be used to place bounds on the space of all unitary CFTs. 

In the numerical bootstrap, the goal is to maximize some physical quantity, such as the first scaling dimension $\Delta_1$, over all unitary solutions to the $P$ equations \eqref{crossf}. In practice, the optimal solution has a finite number of states in the sum. If it happens that the optimal solution has exactly $P/2+1$ states, then it can be found by solving the closed equations \eqref{primal}. We will show that this is the case for 2d modular bootstrap at zero angular potential. 

\bigskip
\noindent This leads to our main analytic result:

\bigskip

\hangindent=0.7cm \textit{In modular bootstrap, at any order $P$ and central charge $c$, there is a solution to the truncated bootstrap equations \eqref{primal}, with scalar gap $\Delta_1^{(c, P)}$, such that all unitary CFTs of central charge $c$ have $\Delta_1 \leq \Delta_1^{(c,P)}$.}

\bigskip

\noindent It follows that general bounds on all unitary CFTs can be found by solving \eqref{primal}. To derive this, we will show that a solution to \eqref{primal} can be used to build an extremal functional, identical to that obtained in the usual numerical bootstrap. Note that this claim does not apply to all solutions of \eqref{primal}, but to one special solution, which must be carefully sought among many.

This is based on the same idea used for numerical bootstrap of 1d correlators in \cite{El-Showk:2016mxr}. Other approaches to the correlator bootstrap using truncated crossing equations have also been discussed in \cite{Gliozzi:2013ysa,Gliozzi:2014jsa,Li:2017ukc}.

\subsection{Comments on Monotonicity}\label{ss:monotonicity}

The largest dimension participating in \eqref{primal}, $\Delta_{P/2}$,  grows with $P$. Therefore the truncation order $P$ acts as a UV cutoff. Qualitatively, there is a parallel between changing $P \to P-2$ and a renormalization group transformation: As $P$ is decreased, the effect of high-dimension operators must be absorbed into the low-dimension spectrum and OPE coefficients, in order to preserve the crossing equations. A corollary of the result stated above is monotonicity:

\bigskip

\hangindent=0.7cm \textit{Under reducing the cutoff $P \to P-2$, the gap $\Delta_1^{(c,P)}$ is non-decreasing.}

\bigskip

\noindent Roughly speaking, we can view a solution to \eqref{primal} as a bootstrap version of an effective field theory, in the following sense. For $P \gg 1$, the low-lying operators $\Delta_\mu$ are insensitive to the truncation, and take their physical values for the target CFT. For $\mu \sim P/2$, the operators of dimension $\Delta_\mu$ are not true operators in the full-fledged CFT, but are effective operators in the cutoff theory, at least for the purposes of the crossing equation. Like the higher dimension operators in a Wilsonian effective Lagrangian, they parameterize all of the contributions from the UV, and very good approximations to the low energy physics can be obtained with a finite number of effective operators. This point of view is useful to keep in mind when interpreting solutions to the truncated equations, because we should not expect operators with $\mu \sim P/2$ to have their physical scaling dimensions; they must also encode the effects of the UV states that are removed by the truncation. We will see this effect in the numerics.

\section{Dual Bootstrap}\label{s:dual}

In this section, we derive the claims of section \ref{s:primal}, by relating the primal equations to the functional bootstrap. We will first recast the functional bootstrap as a nonlinear optimization problem, then dualize to the primal equations.

\subsection{Review}
The functional method \cite{Rattazzi:2008pe} relies on the following observation. Consider the action of a linear functional on $W_{\Delta}$,
\be\label{funalpha}
f(\Delta) = \sum_{i=1}^P \alpha_i f_i(\Delta) \ .
\ee
The $\alpha_i$ parameterize the functional. If we find a functional that is positive on the vacuum, and on all dimensions above some gap,
\be\label{fineq}
f(0) > 0  \quad \mbox{\ and\ } \quad 
f(\Delta) > 0 \mbox{\ for\ } \Delta> \Delta_1
\ee
then every term in the sum \eqref{cross1} is positive for dimensions in this range. Therefore, to satisfy crossing, every unitary CFT must have an operator at or below $\Delta_1$. 

The strongest bounds come from minimizing $\Delta_1$. For our purposes, a useful way to state the optimization procedure is as follows. Fix $P$, the gap $\Delta_1$, and the normalization condition $\alpha_P=1$, and:\footnote{Generally, the optimum may occur either for $\alpha_P>0$ or $\alpha_P<0$. We choose the overall sign of the functionals such that it is always positive.}
\begin{align}\label{lpa}
&\mbox{maximize\ } f(0) \mbox{\ over\ } \alpha_a \\
&\mbox{subject to\ } f(\Delta) \geq 0 \mbox{\ for\ } \Delta \geq \Delta_1 \ . \notag
\end{align}
Here and below, the indices run over the ranges
\begin{align}
a,b,c,\dots  &= 1 \dots P-1 \ , \\
i,j,k,\dots &= 1 \dots P \ .\notag
\end{align}
\eqref{lpa} is a linear program with an infinite number of constraints, or can be recast as a semidefinite program \cite{Poland:2011ey}.  If the constraints are feasible, and the objective $f(0)$ is positive at the maximum, then the initial choice of $\Delta_1$ is ruled out. This procedure is repeated, for different values of $\Delta_1$, until the theory is marginally excluded, \ie $f(0) = 0$ (to some desired precision). This is the edge of the exclusion region, and the optimal functional at this point is called the extremal functional. The \textit{dual bootstrap}
 refers to the problem of finding the extremal functional.

\subsection{Functionals parameterized by zeroes}

\begin{figure}[t]
\begin{center}
\includegraphics[scale=1.0]{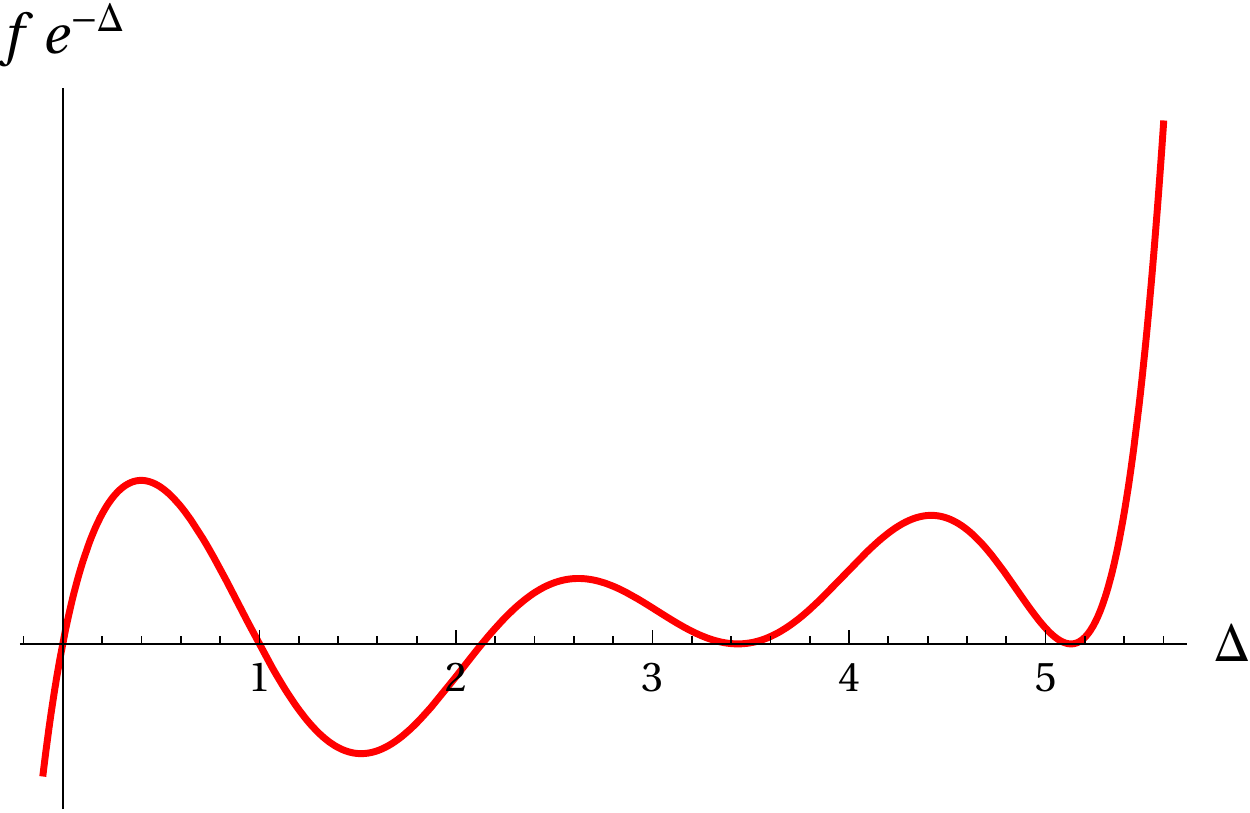}
\end{center}
\caption{\small Example of an extremal functional for modular bootstrap, with $c=12$ and truncation order $P=6$. There are single zeroes at $\Delta_0 \approx 0$ and $\Delta_1 \approx 2.13$, and double zeroes at $\Delta_{2}\approx 3.43$ and $\Delta_3 \approx 5.13$. The additional single zero near $\Delta \approx 1$ plays no role in the discussion. (The root near the origin is slightly shifted due to null state contributions described in section \ref{ss:modularreview}, but this too small to be visible in the plot.)}\label{fig:examplefunctional}
\end{figure}

The zeroes of the extremal functional correspond to solutions of the primal equations \eqref{crossf}. This is a general fact of linear programs, which has been used to find the spectrum of various CFTs, including the critical Ising model in two \cite{ElShowk:2012hu} and three \cite{El-Showk:2014dwa} dimensions.

In certain spinless bootstrap problems, including modular bootstrap, the extremal functional has a very simple pattern of zeroes. Empirically, for even $P$, there are single zeroes at $\Delta = 0$ and $\Delta = \Delta_1$, and double zeroes at $P/2 - 1$ additional states. An example is plotted in figure \ref{fig:examplefunctional}. This set of $P/2+1$ zeroes corresponds to a solution of the truncated primal equations \eqref{primal}.

We will make an ansatz for the functional by assuming that this pattern of zeroes continues at higher $P$. We will confirm \textit{a posteriori} that this assumption holds for modular bootstrap, at least for all $c$ and $P$ that we consider.

Given this assumption about the pattern of zeroes, we now replace the linear optimization problem \eqref{lpa} by a nonlinear optimization. Choose a spectrum $\Delta_\mu$ for $\mu = 1\dots P/2$. From this spectrum, construct the unique functional obeying
\begin{align}\label{funzeroes}
f(\Delta_1) &= 0 \ , \\
f(\Delta_\mu) &=  f'(\Delta_\mu) = 0  \quad (\mu = 2\dots P/2) \ .\notag 
\end{align}
This is $P-1$ linear equations for 
the $P-1$ variables $\alpha_a$, so the solution is unique. In other words, we can view the functional as parameterized by its roots. Next, the linear optimization \eqref{lpa} is replaced by: 
\be\label{newmax}
\mbox{maximize\  } f(0) \mbox{\ over\ } \Delta_\mu, \mbox{\ with\ } \alpha_a
\mbox{\ determined by \  \eqref{funzeroes}} \ .
\ee
The optimization step can also be eliminated altogether by solving directly for the extremal functional. This has $f(0) = 0$ and is an extremum with respect to $\Delta_\mu$. So we simply append to \eqref{funzeroes} the equations 
\be\label{optzero}
f(0) = 0 \ , \qquad \frac{d}{d \Delta_\mu} f(0) = 0 \ ,
\ee
where in the latter equation, $\mu = 2\dots P/2$, and the gradient is taken along solutions to \eqref{funzeroes}. That is, $\alpha_a = \alpha_a(\Delta_\mu)$ as determined by \eqref{funzeroes}.

This is now a closed system of equations for the extremal functional. There is no guarantee that the positivity conditions will be obeyed in this formulation, but this is easily checked after the functional has been constructed, and holds in the applications we will consider. A similar system of equations, but for the linear version of the problem and involving also the primal variables, appears in \cite{El-Showk:2016mxr}. 

To summarize, we have formulated the problem of finding the extremal functional as $3P/2-1$ equations \eqref{funzeroes} and \eqref{optzero} for the $P/2$ dimensions $\Delta_\mu$ and $P-1$ functional parameters $\alpha_a$. In the derivative basis, the equations are polynomial (after absorbing a positive $\Delta$-dependent factor into the coefficients $a_\mu$).

This reduces the problem of finding the extremal functional to the problem of finding zeroes of a system of polynomial equations. The degree of the polynomials grows with $P$, so this is non-trivial. Moreover, we have replaced a convex optimization by a non-convex root-finding problem, where there may be runaways and spurious solutions. In practice, these obstacles can be overcome if we can generate a good guess to use in the first step of Newton's method. We will describe how to do this for modular bootstrap in section \ref{s:newmodular}.

\subsection{Duality}\label{s:duality}
As the nomenclature suggests, the truncated versions of the primal and dual bootstrap are, in fact, Lagrange duals. With our nonlinear formulation of the dual problem, this duality holds under the assumption that the double roots $\Delta_\mu$ of the extremal functional have multiplicity exactly two -- not higher. 

The duality immediately proves the main results stated in section \ref{s:primal}. In the dual bootstrap, monotonicity of the gap under changing $P$ is obvious --- if we allow a more general functional, the bound on $\Delta_1$ can only get stronger. When translated into primal language, this monotonicity property becomes more surprising.

The duality that we describe is already very well known in the linear formulation of the problem. We will rederive it in terms of the nonlinear version above.

We will start with the dual bootstrap --- \ie the problem of finding extremal functionals \eqref{newmax} --- then dualize to the primal bootstrap \eqref{primal}. The constrained optimization \eqref{newmax} is equivalent to maximizing the Lagrangian
\be
L = \alpha_i f_i (0) + a_\mu \alpha_i f_{i}(\Delta_\mu) + \lambda_\nu \alpha_i f_i'(\Delta_\nu)
\ee
over the independent variables $\alpha_a, \Delta_\mu$ and the Lagrange multipliers $a_\mu, \lambda_\nu$ (with $\mu = 1\dots P/2$, $\nu = 2\dots P/2$.) Using the normalization condition $\alpha_P=1$ and rearranging terms,
\be\label{newL}
L = E_P + \alpha_a E_a
\ee
with
\be
E_i \equiv f_i(0) + a_\mu f_{i}(\Delta_\mu) + \lambda_\nu f'_{i}(\Delta_\nu) \ .
\ee
To dualize, we reinterpret this Lagrangian, viewing $\Delta_\mu, a_\mu, \lambda_\nu$ as independent variables, and $\alpha_a$ as a Lagrange multiplier. Assuming $f_i''(\Delta_\mu) \neq 0$ so that no roots have multiplicity higher than two, the equation of motion for $\Delta_\mu$ imposes $\lambda_\nu = 0$. Therefore, the primal optimization problem is
\begin{align}\label{primalopt}
&\mbox{maximize\ } f_P(0) + a_\mu f_P(\Delta_\mu)  \mbox{\ over \ } \Delta_\mu, a_\mu \\
&\mbox{subject to\ }f_a(0) + a_\mu f_a(\Delta_\mu) = 0 \ .\notag
\end{align}
These are the same sums as appear in the primal crossing equation \eqref{primal}, so the constraints consist of the first $P-1$ primal equations. Finally, let us also impose extremality, $f(0) = 0$, so the assumed gap is marginally excluded. This is equivalent to the $P$th primal equation.

The conclusion is that the problem of finding the extremal functional is precisely dual to the problem of solving the $P$ truncated crossing equations, \eqref{primal}, for the spectrum and coefficients. In appendix \ref{s:directduality}, we derive the same statement by direct analysis of the extremal functional equations \eqref{funzeroes}-\eqref{optzero}, and in the process, give a formula for the truncated OPE coefficients $a_\mu$ in terms of the extremal functional.

\subsection{From primal solutions to extremal functionals}\label{ss:getfunctional}

Given a solution to the primal equations \eqref{primal} for some $P$, it is straightforward (and computationally trivial) to construct the dual extremal functional:  Fix $\Delta_\mu$ to the primal values, and solve the linear equations \eqref{funzeroes} for the functional $\alpha_a$. A consistency check is that this functional should vanish on the identity, $f(0) = 0$.

Once a candidate extremal functional has been obtained this way, it can be plotted to confirm the positivity conditions \eqref{fineq}.  If \eqref{fineq} is satisfied, then the functional implies rigorous bounds, so in the end, our bounds are not conditioned on any assumptions about the pattern of zeroes. 

It is also guaranteed that the bounds we derive are optimal, in the sense that linear programming at the same value of $P$ will always produce exactly the same bounds.  It is impossible to derive a stronger bound (at a given $P$), because doing so would contradict the primal solution that we have found. In other words, the method we have described produces solutions which are primal-dual feasible, and therefore optimal.

\section{Modular Bootstrap Algorithm}\label{s:newmodular}

So far, our discussion has applied to a general class of spinless bootstrap problems, with a certain pattern of zeroes in the extremal functional. We now focus on the modular bootstrap exclusively.  In this section we first review the (spinless) modular bootstrap \cite{Hellerman:2009bu,Friedan:2013cba}, then describe our algorithm in detail.

\subsection{Setup}\label{ss:modularreview}
Consider the partition function of a 2d CFT at zero angular potential, 
\begin{align}
Z(\beta) &= \sum_{\rm{\scriptsize states}} e^{-2\pi \beta(\Delta - \frac{c}{12})}\\
&= \sum_{\rm{\scriptsize primaries}} \chi_{\Delta}(\beta)
\end{align}
with $\beta>0$, and $c$ the central charge. In the second line we have organized the sum into characters of the algebra, Virasoro $\times$ Virasoro.\footnote{
$\chi_\Delta(\beta) \equiv \left( \chi^{Vir}_{\Delta/2}(i\beta)\right)^2$, with $\chi_h^{Vir}(\tau)$ the standard chiral Virasoro character. 
} Crossing symmetry in this context is modular invariance: $Z(\beta) = Z(1/\beta)$.  It is convenient to define the reduced partition function 
\begin{align}
\hat{Z}(\beta) &= |\eta(i\beta)|^2 |i\beta|^{1/2}Z(\beta) \label{zzhat}\\
&= \sum_{\mu=0}^\infty \rho_\mu G_{\Delta_\mu}(\beta) \ .\label{zhatss}
\end{align}
where $\rho_\mu$ is the degeneracy of primaries with dimension $\Delta_{\mu}$. This function is also modular invariant, $\hat{Z}(\beta) = \hat{Z}(1/\beta)$. The vacuum is $\Delta_0 = 0$, $\rho_0 = 1$. The Dedekind $\eta$-functions cancel similar factors in $\chi_\Delta$, such that in \eqref{zhatss}, the blocks for non-vacuum states are
\begin{align}
G_\Delta(\beta) &= \beta^{1/2} \exp\left[-2\pi \beta(\Delta - \frac{c-1}{12} ) \right] \ ,
\end{align}
and, for the vacuum,
\be\label{vacblock}
G_{0}(\beta) = \beta^{1/2} e^{2\pi \beta \frac{c-1}{12}}(1-e^{-2\pi \beta})^2 \ .
\ee
(The extra factors in the vacuum block are to account for the null state at level one, $L_{-1}|0\rangle = 0$.)
The crossing equation \eqref{cross1} is
\begin{align}\label{crossZ}
\sum_{\mu=0}^{\infty}\rho_\mu W_{\Delta_\mu}(\beta) &= 0 \\
W_{\Delta}(\beta) &\equiv G_{\Delta}(\beta)  - G_{\Delta}(\frac{1}{\beta})\ .\notag
\end{align}
We choose the basis of derivative functionals
\be\label{calfk}
{\cal F}_k = \frac{1}{2(2k-1)!}\left[\half(1 + \beta)^2\p_\beta\right]^{2k-1} \left.\frac{1 + \beta}{2\sqrt{\beta}}  \right|_{\beta=1} \ ,
\ee
for $k = 1\dots P$. This is a convenient basis because the resulting $f_k$ are simply odd-index Laguerre polynomials, $L_{2k-1}$. Indeed, acting with ${\cal F}_k$ on the crossing equation \eqref{crossZ} gives\footnote{
\textit{Derivation:} The generating function for Laguerre polynomials is
\be
\sum_{p \geq 0}t^p L_p (2w)e^{-w} = (1-t)^{-1} \exp\left( - \frac{1+t}{1-t} w\right) 
\ee
Therefore acting with $\frac{1}{p!} (\p_t)^p|_{t=0}$ on the right-hand side gives $L_p(2w) e^{-w}$. With the identification $\beta = \frac{1+t}{1-t}$, this is the same as ${\cal F}_k[W_{\Delta}(\beta)]$. The even-index Laguerre's do not appear because the corresponding differential operator vanishes on the modular-odd function $W_{\Delta}(\beta)$.
}
\be\label{afzsum}
\sum_{\mu=0}^\infty a_\mu f_k(\Delta_\mu) = 0 \ ,
\ee
where
\be
a_\mu \equiv \rho_\mu e^{-2\pi \Delta_\mu} \ ,
\ee
and, for $\Delta > 0$,
\be\label{fkdef}
f_k(\Delta)  = L_{2k-1}(4\pi x)  \ , \quad x \equiv \Delta- \frac{c-1}{12}  \ .
\ee
The vacuum state is different due to the extra factors in \eqref{vacblock},
\be
f_k(0)  = L_{2k-1}(4\pi x_0)  - 2 e^{-2\pi} L_{2k-1}(4\pi(x_0+1)) + e^{-4\pi} L_{2k-1}(4\pi(x_0+2)) \ ,
\ee
with $x_0  = -\frac{c-1}{12}$. 

At this point, we have written the primal bootstrap equations in the same format as section \ref{s:primal}. The truncated version, a closed system of $P$ equations for $P$ unknowns, is
\be\label{truncmod}
\sum_{\mu=0}^{P/2} a_\mu f_k(\Delta_\mu) = 0 \ ,
\ee

\subsection{Algorithm}\label{ss:algorithm}

We now turn to numerical algorithms based on the reformulation of the crossing problem described in sections \ref{s:primal}-\ref{s:dual}. We focus on the simplest bootstrap question, which is to place an upper bound on the scalar gap $\Delta_1$, given the central charge $c$. The algorithms can be modified to accommodate similar problems.

We can consider two main approaches:
\begin{itemize}
\item \textit{Dual}: Solve the functional equations \eqref{funzeroes} and \eqref{optzero}, for the spectrum and the functional.
\item \textit{Primal}: Solve the $P$ truncated primal crossing equations, \eqref{primal}, for the spectrum and degeneracies.
\end{itemize}
Note that standard optimization methods for functional bootstrap require an additional scan over $\Delta_1$. In our case, $\Delta_1$ is a free parameter, so this step is not necessary.

For modular bootstrap, both the dual and primal algorithms are much faster than semidefinite programming. In the rest of this paper we will consider only the primal algorithm, which was somewhat faster in our preliminary testing, and conceptually simpler. 

The purpose of the algorithm is to solve the equations \eqref{truncmod} for some large value of $P$. The strategy is to start at small $P$, solve the equations by Newton's method, then gradually increase $P$, using the previous results to generate a good initial guess for Newton's method. The same strategy was used for 1d correlator bootstrap in \cite{El-Showk:2016mxr} (specifically, our algorithm is similar to the `upgrading + error correction' method). Besides the differences between 1d correlator bootstrap and 2d modular bootstrap, our algorithm has two significant improvements:  better numerical stability, and a more elaborate, more accurate method to guess the initial points. This will allows us to increase $P$ in large increments.

For numerical stability, we first introduce normalization factors to rescale the $P \times P/2$ matrix $f_k(\Delta_\mu)$. Rewrite \eqref{truncmod} as
\be\label{rescalemod}
\sum_{\mu=0}^{P/2}\tilde{a}_\mu \tilde{f}_k(\Delta_\mu) = 0 \ ,
\ee
where
\begin{align}
\tilde{a}_\mu &= a_\mu |\max_k f_k(\Delta_\mu)| \\
\tilde{f}_k(\Delta_\mu) &= \frac{f_k(\Delta_\mu)}{f_k(0) |\max_k f_k(\Delta_\mu)| } \ .
\end{align}
The matrix $\tilde{f}_k(\Delta_\mu)$ is scaled such that the vacuum column is all 1's, and every other column has an absolute maximum entry of $\pm 1$.

 We now apply Newton's method to the equations \eqref{rescalemod} for the $P$ unknowns $(\tilde{a}_\mu, \Delta_\mu)$. The Laguerre polynomials, and their derivatives, are calculated by 3-term recurrence relations. This is faster and more numerically stable than direct evaluation. Our implementation of Newton's method is completely standard so we will not review it. 
 
An important element in the algorithm is a guess generator, to provide the initial point in Newton's method.  The method is largely insensitive to the initial guesses for $\tilde{a}_\mu$, so for these coefficients we always guess $\tilde{a}_\mu = 1$. Convergence is very sensitive to the starting $\Delta_\mu$, so these must be chosen carefully. We start by solving the equations for $P=4,6$. For these initial points, it is easy to find a good guess by hand (or using a linear/semidefinite optimizer such as SDPB \cite{Simmons-Duffin:2015qma}). Then we proceed upwards in $P$ recursively, using previous results to generate a guess for $\Delta_\mu$.
 
\begin{figure}[t]
\begin{center}
\hspace{-.85cm}\includegraphics[scale=0.71]{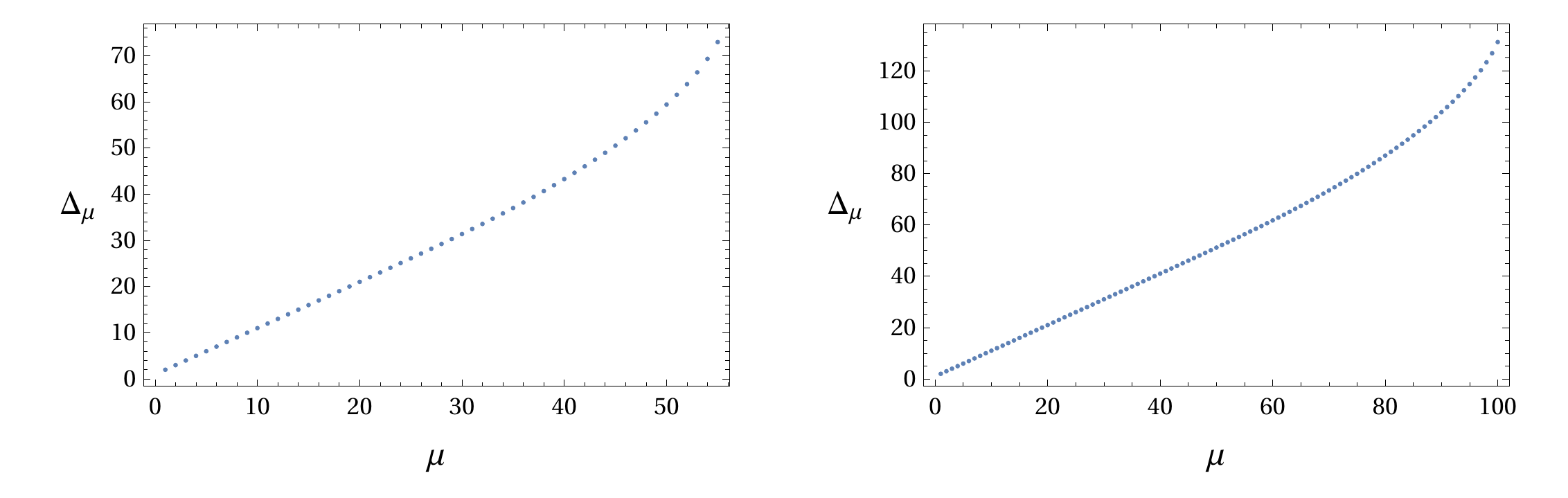}
\end{center}
\caption{\small
Spectrum of the $c=12$ solution at truncation order $P=110$ (left) and $P=200$ (right). The horizontal axis is the state number, $\mu = 1, 2,\dots,P/2$. Note that the curves are approximately the same shape. This observation is used to generate very accurate guesses for the initial point in Newton's method.
\label{fig:selfsim}}
\end{figure}

 Our guess generator is based on the observation that the  spectrum, at different values of $P$, is self-similar. That is, the curve $\Delta_\mu$ as a (discrete) function of $\mu$ has almost exactly the same shape for all $P$, once $P$ is large enough. This is illustrated by the plots in fig.~\ref{fig:selfsim}, which are the exact spectra at $P  = 110$ and $P  = 200$ for the $c=12$ modular bootstrap. The shapes in the two plots are almost identical; only the scaling of the axes and the state spacing differs. This motivates a guess where we rescale the lower-$P$ spectrum, and re-sample at the appropriate points. This generates a good guess, and we use precisely this method for $P \lesssim 200$.  However, we can do better, and take larger $P$-steps, by correcting for the fact that the spectrum is not quite self-similar. We do this by parameterizing the shape curve, fitting the parameters as functions of $P$, and extrapolating. The full algorithm is given in appendix \ref{app:guessgenerator}.

 The resulting guesses are very accurate. For example, in the $c=100$ modular bootstrap, using data from $P=120,122,124,\dots,200$, the extrapolation to $P = 280$ produces a spectrum $\Delta_{\mu}^{\rm{guess}}$ that differs from the exact spectrum by under one part in $10^3$-$10^5$. Results at higher $P$ are comparable. This means that we never need to run more than a handful of Newton steps to converge to an accurate spectrum.

This completes the algorithm. The only computationally expensive step is inverting the Jacobian in Newton's method, which scales roughly as $P^3$. The computations are performed in Mathematica with high precision arithmetic.

\section{Modular Bootstrap Results}\label{s:results}

\subsection{3d gravity and summary of existing bounds}

Three-dimensional gravity in the semiclassical limit is holographically dual to a 2d CFT with a large central charge, $c \gg 1$. Quantum fields in the bulk, aside from the graviton, are dual to low-dimension operators in the CFT. It follows that the CFT should also have a sparse spectrum of low-dimension operators, in that the number of states below any fixed $\Delta_*$ should be finite as $c \to \infty$.

`Pure' 3d gravity, which consists only of the graviton plus black holes, would have no new primaries between the vacuum and black hole threshold. That is, $\Delta_1 \sim \frac{c}{12}$. It is unknown whether pure gravity exists as a quantum theory. A putative partition function was suggested by Witten \cite{Witten:2007kt}, but explicit calculations by Maloney and Witten \cite{Maloney:2007ud} gave a different answer that was incompatible with a holographic interpretation. It is possible that this calculation can be modified to produce a consistent quantum theory \cite{Keller:2014xba}, but at this point, the realm of possibilities is poorly understood.  

Even the simplest question is unanswered: What is the largest gap $\Delta_1$ compatible with diffeomorphism invariance? Large diffeomorphisms in 3d gravity correspond to modular transformations in the dual CFT, so this is a question for the modular bootstrap.

The functional approach to modular bootstrap was initiated in \cite{Hellerman:2009bu}, where, working analytically at $P=2$, it was proved that $\Delta_1 \leq \frac{c}{6} + 0.474$. This is known as the Hellerman bound. At large $c$, the Hellerman bound is a factor of 2 above the black hole threshold. It has also been shown analytically that, for $\Delta \gtrsim \frac{c}{6}$, 2d CFTs with a large gap have the same entropy as BTZ black holes \cite{Hartman:2014oaa}. However, the range $\frac{c}{12} \leq \Delta \leq \frac{c}{6}$ remains enigmatic. This range certainly does not have a universal spectrum, even in holographic theories, but it is an open question whether there must be any states at all within this window.

Numerical evidence strongly suggests that the asymptotic bootstrap bound can be improved to somewhere between $c/12$ and $c/6$ \cite{Friedan:2013cba,Qualls:2013eha,Collier:2016cls}. (For related work, see \cite{Hellerman:2010qd,Keller:2012mr,Chang:2015qfa,Shaghoulian:2015kta,Kim:2015oca,Benjamin:2016fhe,Das:2017vej,Cho:2017fzo,Dyer:2017rul,Apolo:2017xip,Afkhami-Jeddi:2017idc,Bae:2017kcl,Collier:2017shs,Bae:2018qym,Anous:2018hjh}). For $c \lesssim 150$, numerical bounds are as strong as $\Delta_1 \lesssim c/8.2$. However, $c \sim 150$ is the largest central charge accessible to standard numerical methods, and at this value, it is clear that the bound has not yet converged. Results at larger $c$ are needed to find the bound in the semiclassical, large-$c$ limit.

The difficulty is that for large $c$, the bound on $\Delta_1$ converges  slowly with $P$, so large $c$ requires large $P$. This is a general obstacle to bootstrapping quantum gravity in the regime $\Delta \sim c$, which presents a significant challenge to probing black hole physics with the bootstrap. (On the other hand, many interesting questions in AdS/CFT already arise for states with $\Delta \ll c$, and in these cases numerical bootstrap has been very successful.)

The largest truncation order considered in the literature on modular bootstrap is $P=92$ \cite{Collier:2016cls}, using the semidefinite program solver SDPB \cite{Simmons-Duffin:2015qma}. We apply our algorithm up to $P \sim 2250$ (derivative order 4500). At small $P$, we have checked that our bounds agree exactly with those produced by SDPB.

\subsection{Bound as a function of $c$}

\begin{figure}[t]

\begin{overpic}[scale=1.0]{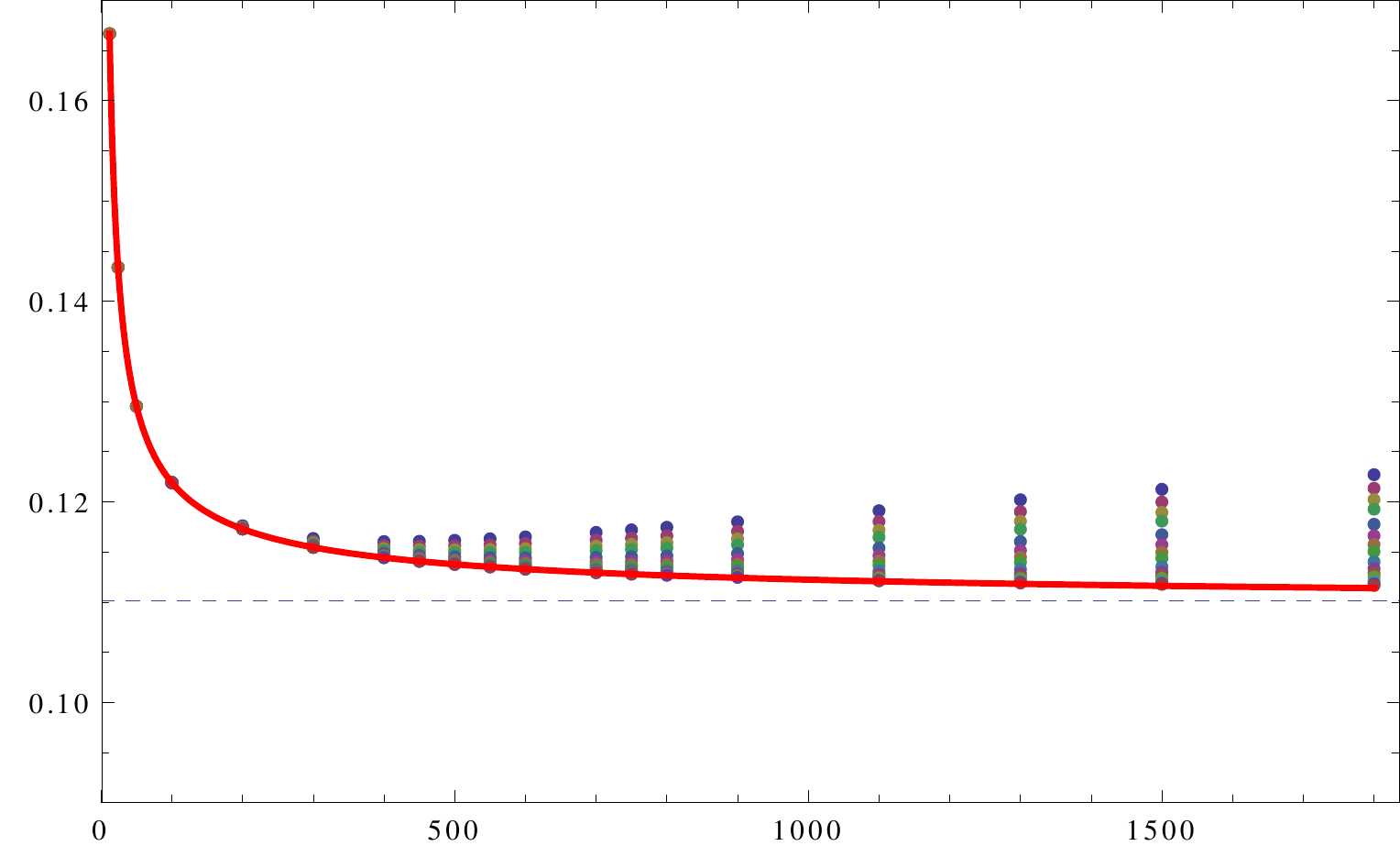}
\put(980,0){$c$}
\put(100,500){$\Delta_1/c$}
\end{overpic}

\caption{\small Upper bound on $\Delta_1/c$, as a function of $c$. Dots are numerical data for truncation at $P = $ 270, 310, 350, 390, 470, 550, 630, 710, 870, 1030, 1350, 1670, 1990, 2310, from top to bottom. The solid red line is the extrapolation to $P = \infty$.  The dashed blue line is the asymptotic estimate at large $c$, $1/9.08$.\label{fig:delta1}}
\end{figure}

The upper bound on $\Delta_1/c$ is plotted in fig.~\ref{fig:delta1}. The dots are numerical data, and the solid line is the extrapolation to $P = \infty$ with an exponential fit, $\Delta_1(P) = \Delta_1^{\infty} + a e^{-b P}$.

To estimate the asymptotic bound, we fit the data in the range $300 \leq c \leq 700$ to a function of $c$. In this range, the bound has converged very close to its $P \to \infty$ limit. The estimate depends on what function is used in the fit, but natural choices all give estimates clustered around the same value,
\be
\Delta_1 \lesssim  c / 9.1 \ .
\ee
Of course this is not definitive, because we are working at finite $c$, but the errors in the fit are very small.  Different choices of fitting function, together with the adjusted $R^2$ of the fit, are reported in table \ref{table:fits}. We also give an error estimate $\delta_c$, which is defined as the fractional error of our two extrapolations. For example, $\delta_{900}$ compares the extrapolation in $c$  (which was found using data only from $c \leq 700$) to our actual data at $c=900$, extrapolated up to $P=\infty$. The errors are smallest for the logarithmic fit,
\be
\Delta_1^{\rm max}  \approx \frac{c}{9.08} + 0.439 \log c  - 0.896 \ .
\ee
The best strict (non-extrapolated) bound on $\Delta_1/c$ is $c=1800, P=2310$, for which $\Delta_1^{\rm max} = c/8.956$.

\begin{table}
\centering
\caption{Fitting the bound as a function of $c$ \label{table:fits}}
\vspace{0.3cm}
\begin{tabular}{ccccc}
\hline\hline
Fitting Function &  $1 - R^2_{\rm{adj}}$ &  $\delta_{900}$ & $\delta_{1800}$ & Best fit $A$ \\
\hline\small
$c/A + B_1$ &  $10^{-7}$ & $9 \times 10^{-4}$ & $3 \times 10^{-3}$ & 9.01\\
$c/A + B_1 + B_2/c$ & $10^{-9}$ & $2 \times 10^{-4}$ & $9 \times 10^{-4}$ & 9.04\\
$c/A + B_1 + B_2 \log c$ & $6 \times 10^{-11}$ &  $4 \times 10^{-6}$ & $2 \times 10^{-5}$  & 9.08\\
\hline\hline
\end{tabular}
\end{table}

\begin{figure}[t]
\begin{center}
\hspace{-.85cm}\includegraphics[scale=0.71]{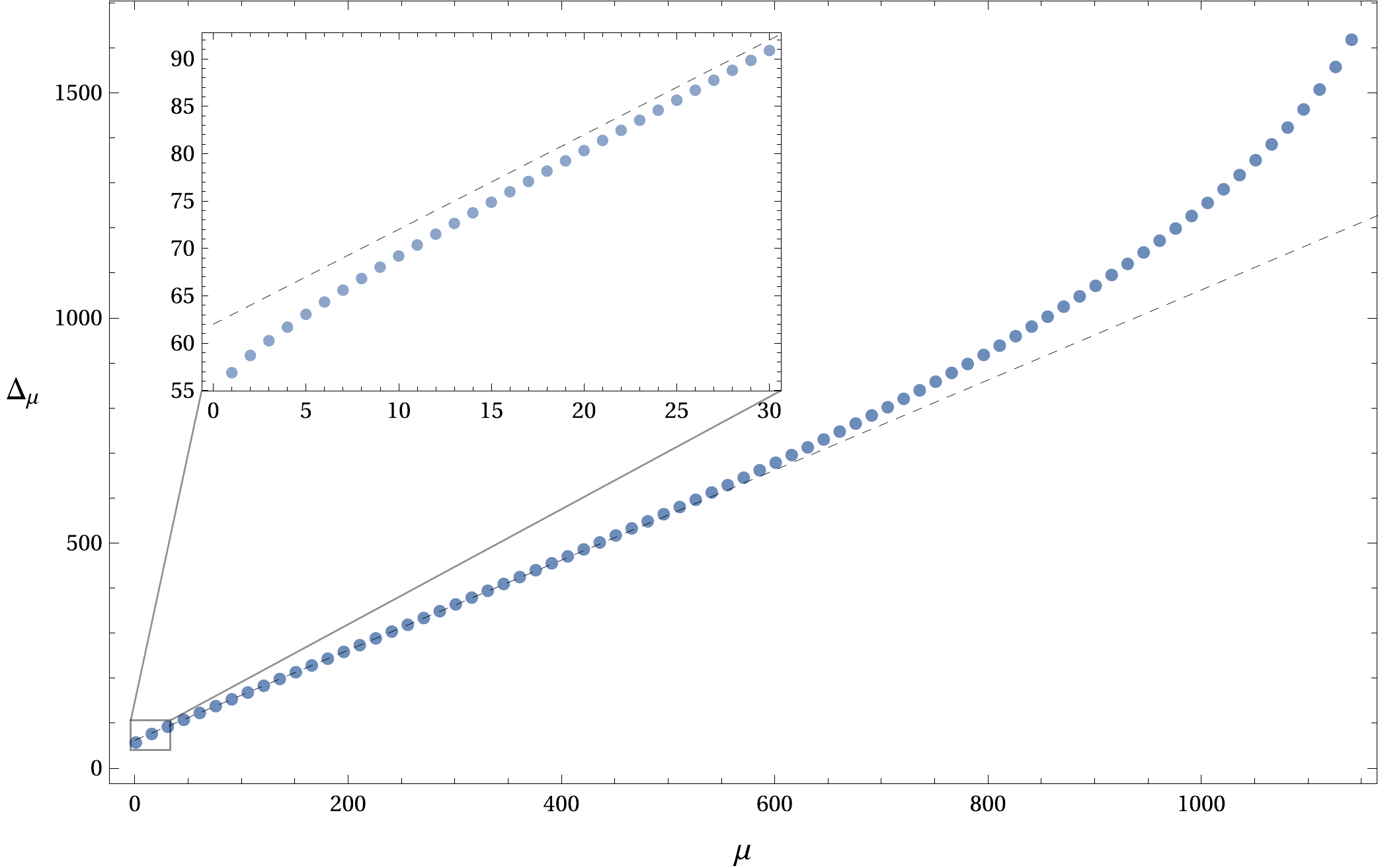}
\end{center}
\caption{\small The numerical spectrum $\Delta_\mu$ for $c=500$ and $P=2310$. In the larger plot we show every 15th scaling dimension to reduce clutter.  The smaller inset figure shows the first 30 scaling dimensions with deviation away from linearity. In \cite{upcoming}, the linear regime of the spectrum with $\Delta_\mu = \frac{c-4}{8} + \mu$ (dashed line) will be explained analytically.
\label{fig:c500inset}}
\end{figure}

In addition to the gap, $\Delta_1$, the method also provides a spectrum with over a thousand operators. As an example, the spectrum for $c=500$, $P = 2310$ is plotted in fig.~\ref{fig:c500inset}. As discussed in section \ref{ss:monotonicity}, only the operators $\mu \ll P$ should be trusted, but in practice, roughly the lower half of the spectrum with $\mu \lesssim P/4$ is converged.

\subsection{$c=12$ and the modular $j$-function}
The case $c=12$ is especially interesting, because the numerics appears to converge to a spectrum with integer scaling dimensions $\Delta_\mu = 2,3,\dots$, and integer degeneracies. (The case $c=4$ has a similar story.) The bound on the gap at $c=12$, obtained with $P=2000$ in a few hours of CPU time, is:
\be
\Delta_1 < 2 + 1\times 10^{-30} \qquad (c=12,\ P=2000) \ .
\ee
The numerical solution of the truncated crossing equations at $P=2000$ is
\begin{align}
\hat{Z} &\approx q^{1/12}\beta^{1/2} \big[q^{-1}(1-q)^2 + (196882 + 2 \times 10^{-24}) q \notag\\
& \qquad + (21099994 + 4 \times 10^{-22}) q^2 + (821115567+3 \times 10^{-20})  q^3 + \cdots \big]
\end{align}
where $q \equiv e^{-2\pi \beta}$. The degeneracies converge to integers that suggest a link to the modular $j$-function. Indeed, reintroducing the $\eta$-functions removed in \eqref{zzhat}, this corresponds to
\be
Z \approx j(\tau) - 744 \ , \qquad \tau = i\beta \ .
\ee
 This coincidence will be explained analytically in \cite{upcoming}. Note that we did not impose invariance under $\tau \to \tau + 1$, so it is surprising to find the $j$-function appearing.

\subsection{Algorithm Benchmarks}

\begin{figure}[t]

\begin{center}
\includegraphics[scale=0.7]{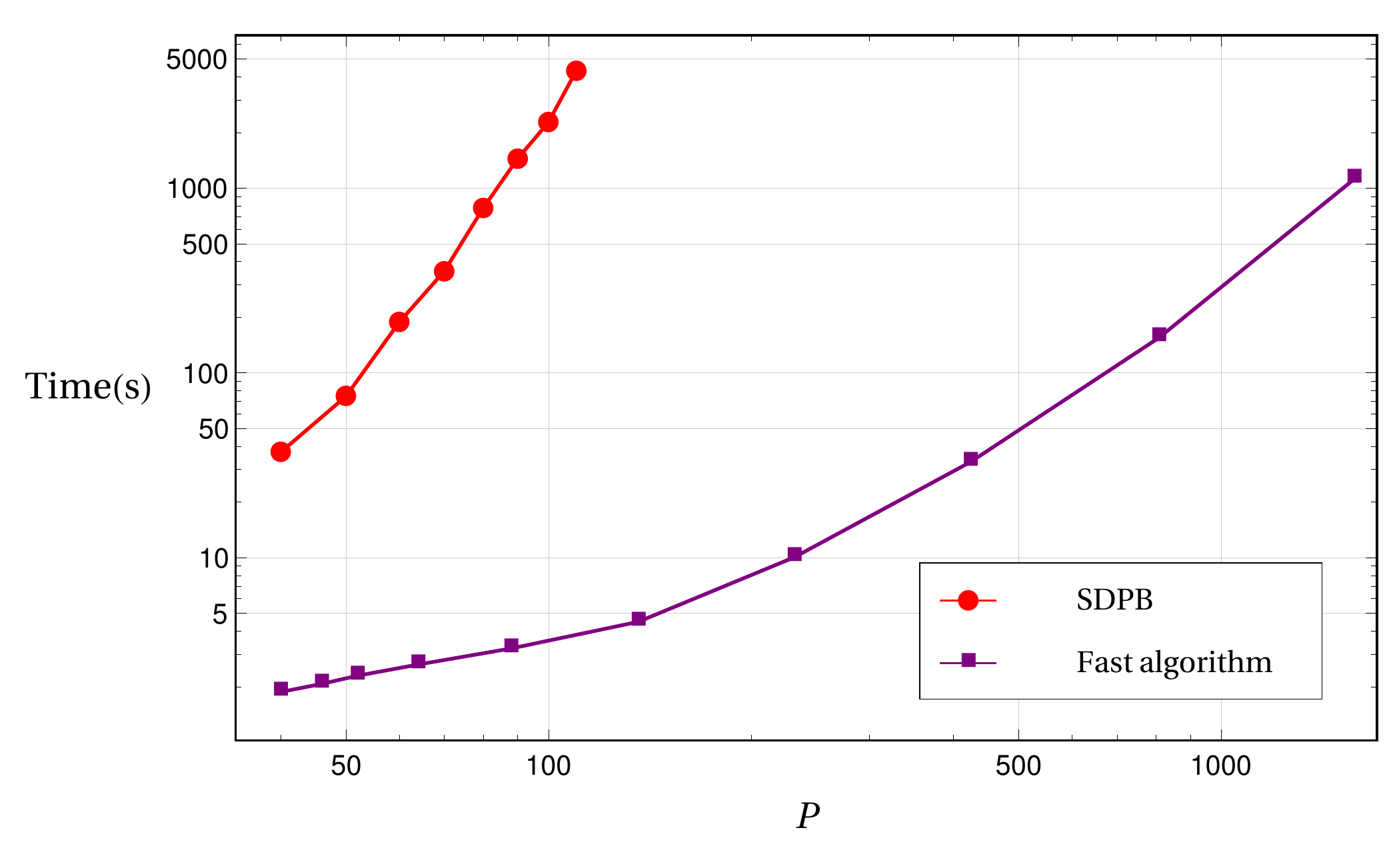}
\end{center}

\caption{\small Runtime of SDPB and our Newton-based algorithm for modular bootstrap at c=12, vs the number of polynomials, $P$. Larger $c$ requires more precision, so runtimes are somewhat longer. \label{fig:runtime}}
\end{figure}

Runtime is compared to semidefinite programming with SDPB (v1) \cite{Simmons-Duffin:2015qma} in fig. \ref{fig:runtime}. Extrapolating the runtime of SDPB to large $P$ may be  unreliable, but naively fitting to a power law and setting $P=2000$ gives the estimate $\sim 10^9$ s.

We can also compare to the extremal functional method of El-Showk and Paulos \cite{El-Showk:2016mxr}.  They solve a different problem (1d correlators) so it is not a direct comparison, but they report runtime of 45 minutes at $P = 150$. For modular bootstrap, our algorithm runs in seconds at the same $P$. The algorithm is very similar to that described in \cite{El-Showk:2016mxr}, so we do not know the exact origin of the speedup, but it presumably comes from some combination of improvements to numerical stability, the guess generator described in section \ref{ss:algorithm}, or inherent differences in modular bootstrap.\footnote{It is also not clear to us whether the calculations in \cite{El-Showk:2016mxr} were done using the primal equations alone, or the primal-dual formulation of the problem. We use only the primal equations.}

Our procedure is numerically stable, in the sense that round-off errors do not accumulate dramatically. This means that the same accuracy as semidefinite programming is achieved with much lower precision at intermediate steps. (This gives a significant speed-up, but is not the main factor.) For example, at $c=12$, machine precision (16 decimal digits) is sufficient up to $ P \sim 90$, and 60 digits is sufficient for $P \sim 2000$. Calculations at larger $c$ require higher precision, because more terms participate significantly in the crossing equations. The data in fig.~\ref{fig:delta1} was produced with a range of precisions from 16 to 500 digits, chosen to ensure that the matrix inversion step in Newton's method is well behaved.

\section{Discussion}\label{s:discussion}
In summary, we have described a reformulation of the modular bootstrap equations, amenable to both analytic and numerical analysis. On the analytic side, the main logic was to show that $(i)$  extremal functionals can be efficiently parameterized by their roots, and $(ii)$ this is dual to solving the truncated crossing equations. 

For modular bootstrap, our algorithm is orders of magnitude faster than semidefinite programming or any other previous method, extending the maximum truncation order from $P\sim 100$ to $P \sim 2500$. Other potential advantages include:
\begin{itemize}
\item It is numerically stable, so intermediate steps do not require very high precision. 
\item The method might apply to non-derivative or non-polynomial functionals. (One choice has been considered  \cite{Echeverri:2016ztu}.) We restricted to derivative functionals only because in this basis, we have confirmed explicitly (in examples) the pattern of zeroes invoked in section \ref{s:dual}. The method applies to any basis of functionals with a similar pattern of zeroes. 
\item The algorithm can be sped up significantly by hot starting, \ie inputting a guess for the spectrum of the target CFT. This means that analytic results can potentially be leveraged for faster numerics, similar to the way we used results at small $P$ for faster convergence at large $P$.
\item Although the strict bounds we derived apply only to unitary CFTs, the truncated primal equations may also have solutions corresponding to non-unitary CFTs, with potential applications to weakly first order phase transitions \cite{Poland:2018epd,Gorbenko:2018ncu}. This would be similar to the approach taken in \cite{Gliozzi:2013ysa,Gliozzi:2014jsa}. 
\item The truncated bootstrap equations \eqref{primal} can be studied analytically. This may lend additional insights.
\end{itemize}

There is also a major drawback to our approach, at least in our current implementation. We considered only bootstrap problems without spin dependence. More generally, for example in the modular bootstrap with angular potential, or the correlator bootstrap of the 3d Ising model, there are more constraints on the extremal functional, labeled by $\ell$. This typically leads to a different pattern of zeros, not the simple pattern we observed here. In primal language, the problem is that with spin, the solution of the primal optimization problem at truncation order $P$ typically does not have exactly $P/2 + 1$ states.\footnote{The simple pattern of zeros in the spinless modular bootstrap may be related to total positivity of the functionals, which was noted in \cite{Arkani-Hamed:2018ign}. Totally positive functionals lead to polynomials $f_i(\Delta)$ with interlacing roots \cite{pinkus1996spectral}. This fact seems to give the polynomials the right shape to be able to produce one new double root with the addition of two new polynomials.}  

This makes it impossible to truncate to a closed, primal-only problem, as we have done here. Instead one can write primal-dual equations for the extremal functional, involving $(\alpha_a, a_\mu, \Delta_\mu)$ simultaneously, as in \cite{El-Showk:2016mxr}. In principle, we could use Newton's method to solve this enlarged problem, but in practice, we have not found any efficient way to generate initial guesses. The pattern of states does not change predictably as the truncation order is increased.

As described in the introduction, part of our motivation in this work was to push the numerical bootstrap to the point where it can be used to construct, or rule out,  candidate theories of 3d gravity. However, our bound $\Delta_1 \lesssim c/9.1$ does not saturate the black hole threshold, so it appears that spinless modular bootstrap is not powerful enough to attack these problems.  It might be possible with spinning modular bootstrap, but spin requires new methods. 

Eventually, it may be possible to use numerics to settle longstanding questions in quantum gravity: Do all theories of quantum gravity require new states well below the Planck scale, such as strings or Kaluza-Klein modes (see e.g. \cite{Belin:2014fna}) ? Does quantum gravity in the ultraviolet rule out a swampland of effective theories in the infrared \cite{Vafa:2005ui}?  Is there a quantum theory of pure gravity in three dimensions \cite{Witten:2007kt}? Versions of all these questions, in anti-de Sitter, can be posed in terms of the space of solutions to the crossing equations. Of course it would be better to solve these problems analytically, but definitive numerical answers would be a good start. 

\bigskip

\bigskip

\textbf{Acknowledgments} We are grateful to David Simmons-Duffin for discussions of semidefinite and simplex methods and to Dalimil Maz\'{a}\v{c} and Leonardo Rastelli for comments on the draft.  We also thank other members of the Simons Bootstrap Collaboration, and the organizers of the bootstrap workshops at ICTS in 2017, and the Azores and Caltech in 2018, as well as the Aspen Center for Physics, where much of this work was done or inspired. This work is supported by Simons Foundation grant 488643.

\bigskip

 \appendix

\section{Direct derivation of optimization duality}\label{s:directduality}
In this section, we re-derive the duality of section \ref{s:duality} by direct analysis of the extrema of the functional optimization \eqref{newmax}. This is useful in practice for translating between dual and primal solutions. Denote 
\be
f_{i\mu} \equiv f_i(\Delta_\mu) \ , \quad f_{i\mu}' \equiv f_i'(\Delta_\mu) , \quad f_{i\mu}'' \equiv f_i''(\Delta_\mu)
\ee
and form the block matrix
\be\label{defv}
V_{i\beta} = [f_{i\mu} \ \ f'_{i\nu}] 
\ee
where $i=1\dots P$, $\mu = 1\dots P/2$, $\nu = 2\dots P/2$, and $\beta=1\dots P-1$. The roots \eqref{funzeroes}, using \eqref{funalpha}, impose
\begin{align}\label{avcon}
\alpha_i V_{i\beta} = \alpha_a V_{a \beta} + V_{P\beta} = 0 \ ,
\end{align}
where we have used the normalization condition $\alpha_P=1$. Therefore, given $\Delta_{\mu}$, the functional with the correct zeroes is
\be
\alpha_a = - V_{P\beta} V^{-1}_{\beta a} \ ,
\ee
where $V^{-1}$ denotes the inverse of the square submatrix $V_{a\beta}$ (recall $a$ is restricted to the range $1\dots P-1$). The functional evaluated on the identity term is
\be
f(0) =\alpha_i f_i(0)  = f_P(0) - V_{P\beta} V^{-1}_{\beta a} f_a(0) \ .
\ee
According to \eqref{newmax}, we seek to maximize this over $\Delta_\mu$. Ultimately we are interested in a maximum with $f(0)=0$, so we can assume the maximum is actually obtained, and therefore occurs at a critical point. The gradient, with $\del_\mu \equiv \frac{d}{d \Delta_{\mu}}$, is
\begin{align}
\del_{\mu} f(0) &= 
- (\del_\mu V_{P\beta}) V^{-1}_{\beta a}f_a(0)
 + V_{P\beta} V^{-1}_{\beta b} (\del_\mu V_{b \gamma} ) V^{-1}_{\gamma a} f_a(0)
\\
&= - \alpha_i (\del_\mu V_{i\beta}) V^{-1}_{\beta a}f_a(0)\\
&= -\alpha_i f_{i\mu}' V^{-1}_{\mu a} f_a(0) - \alpha_i f''_{i\mu} V^{-1}_{\mu + P/2-1,a}f_a(0) \\
&= - \alpha_i f_{i\mu}'' V^{-1}_{\mu + P/2-1,a}f_a(0)
\end{align}
Here $\mu =2 \dots P/2$  is not summed. The first line uses the matrix identity $\del X^{-1} = - X^{-1} \del X X^{-1}$; the second line repackages the two terms into $i = 1\dots P$; the third line uses \eqref{defv}; and the last line follows from \eqref{avcon}.

Therefore we have two options: $f_{i\mu}'' = 0$, so some $\Delta_\mu$ is actually a triple rather than double root, or 
\be\label{vinv}
V_{\beta a}^{-1}f_a(0) = 0 \qquad (\beta = P/2+1 \dots P-1)
\ee
 We assume the latter, so that no roots have multiplicity higher than two. To understand the resulting constraint, define $c_\beta = V^{-1}_{\beta a}f_a(0)$. \eqref{vinv} states that the vector $c_\beta$ takes the block form
 \be
 c_\beta = [-a_\mu \ \ 0_\nu] \ ,
 \ee
 where $\mu=1\dots P/2$ and $0_\nu$ is a vector of $ P/2-1$ zeros. Inverting gives
 \be
 f_a(0) = -  a_{\mu} f_{a \mu}  \ .
 \ee
 Therefore, letting $a_0 = 1$ and $\Delta_0 = 0$, we have
 \be
 \sum_{\mu = 0}^{P/2} a_\mu f_a(\Delta_\mu)  = 0 \ .
 \ee
This is precisely the first $P-1$ equations of the primal bootstrap, \eqref{primal}, with $a_\mu$ the OPE coefficient. Now let us impose extremality, \ie that we are at the boundary of the excluded parameter space. That is, we let $\Delta_1$ be an independent variable and impose
\be
0 = f(0) = \alpha_a f_a(0) + f_P(0)  = -V_{P\beta}V^{-1}_{\beta a} f_a(0) + f_P(0) = -V_{P\beta}c_\beta + f_P(0) = V_{P \mu} a_\mu + f_P(0) \ .
\ee
This is the $P$th equation of \eqref{primal}.

Therefore, the solution of \eqref{newmax}, at extremality, is equivalent to a solution of the $P$ primal bootstrap equations \eqref{primal} (assuming no roots of multiplicity greater than two). We have also shown that, given the extremal functional, the OPE coefficients in the truncated primal solution are
\be
a_\mu = -V^{-1}_{\mu a}f_a(0) \ . 
\ee

\section{Generating guesses for Newton's method}\label{app:guessgenerator}

The logic of our guess generator is described in section \ref{ss:algorithm}.  We  fit to the curves in fig.~\ref{fig:selfsim}, extrapolate in $P$ to generate a new curve, then sample at the appropriate discrete points. In this appendix we describe the complete algorithm. The choice of parameters and fitting functions that we use is ad hoc, but produces good results.   

 Suppose we have already found the spectrum for various truncations $P=P_A$, $A=1,\dots, K-1$. Denote these spectra by $\Delta_\mu^{P_A}$. Define $\Delta^{P_A}(x)$ to be the continuous function that numerically interpolates this spectrum at integer $x$, \ie
 \begin{align}
\Delta^{P_A}(\mu) &= \Delta_\mu^{P_A}  \quad \mbox{for} \quad \mu=1,2,\dots,P_A/2 \ .
\end{align}
We use a degree-22 Hermite interpolation (in Mathematica: {\tt Interpolation[\dots, InterpolationOrder->22]}). Now suppose we want to guess the spectrum at truncation order $P_K$. Let
\be
F_\mu^{P_A} = \Delta^{P_A}\left( \frac{\mu P_A}{P_K} \right) \ .
\ee
This is defined at discrete $P_A$. Fit this this, as a function of $P_A$, to the continuous function
\be
F_\mu(y) = a_1 + a_2 y + a_3 \log(y) + a_4 \log(y)^2 + a_5 \log(y)^3 \ .
\ee
The parameters $a_i$ are chosen such that $F_\mu(P_A) \approx F_\mu^{P_A}$, with least-squares fitting. Finally, our guess for the spectrum at $P = P_K$ is
\be
\Delta_\mu^{\rm{guess}} = F_\mu(P_K) \ .
\ee 

\section{High precision bounds with linear programming}
\label{app:secant}

In this appendix we describe a simple method to find high-precision bounds on the gap, $\Delta_1$, with linear programming. This is used when we compare our results to SDPB, but it is more general. For example, it is faster than the usual method to find bounds on the 3d Ising model (the `kink' in \cite{ElShowk:2012ht}), if the bounds are needed to high accuracy.

The standard method is to run a program solver such as SDPB in feasibility mode, to rule in or out each gap $\Delta_1$, and to zoom in on the optimal $\Delta_1$ using bisection. Instead, we run the linear program \eqref{lpa}. Call the objective at the optimum $M(\Delta_1)$. To find the optimal bound, \ie the gap $\Delta_1$ which is marginally excluded, we use the secant method to solve $M(\Delta_1)=0$.

In the first few steps, this method is slower, because the program solver takes longer to solve an optimization problem than a feasibility problem. But the secant method converges exponentially faster than bisection, so for high accuracy bounds on $\Delta_1$, this method is faster.

\end{spacing}
\small
\bibliography{fast}

\providecommand{\href}[2]{#2}\begingroup\raggedright\begin{thebibliography}{10}

\bibitem{Belavin:1984vu}
A.~A. Belavin, A.~M. Polyakov and A.~B. Zamolodchikov, {{Infinite Conformal
  Symmetry in Two-Dimensional Quantum Field Theory}},
  \href{http://dx.doi.org/10.1016/0550-3213(84)90052-X}{Nucl. Phys. {\bf B241},
  333--380, 1984}.

\bibitem{Poland:2018epd}
D.~Poland, S.~Rychkov and A.~Vichi, {{The Conformal Bootstrap: Theory,
  Numerical Techniques, and Applications}},
  \href{http://dx.doi.org/10.1103/RevModPhys.91.015002}{Rev. Mod. Phys. {\bf
  91}, 15002, 2019},
  [\href{http://arxiv.org/abs/arXiv:1805.04405}{{arXiv:1805.04405 [hep-th]}}].

\bibitem{Rattazzi:2008pe}
R.~Rattazzi, V.~S. Rychkov, E.~Tonni and A.~Vichi, {{Bounding scalar operator
  dimensions in 4D CFT}},
  \href{http://dx.doi.org/10.1088/1126-6708/2008/12/031}{JHEP {\bf 12}, 031,
  2008}, [\href{http://arxiv.org/abs/arXiv:0807.0004}{{arXiv:0807.0004
  [hep-th]}}].

\bibitem{Poland:2011ey}
D.~Poland, D.~Simmons-Duffin and A.~Vichi, {{Carving Out the Space of 4D
  CFTs}}, \href{http://dx.doi.org/10.1007/JHEP05(2012)110}{JHEP {\bf 05}, 110,
  2012}, [\href{http://arxiv.org/abs/arXiv:1109.5176}{{arXiv:1109.5176
  [hep-th]}}].

\bibitem{ElShowk:2012ht}
S.~El-Showk, M.~F. Paulos, D.~Poland, S.~Rychkov, D.~Simmons-Duffin and
  A.~Vichi, {{Solving the 3D Ising Model with the Conformal Bootstrap}},
  \href{http://dx.doi.org/10.1103/PhysRevD.86.025022}{Phys. Rev. {\bf D86},
  025022, 2012}, [\href{http://arxiv.org/abs/arXiv:1203.6064}{{arXiv:1203.6064
  [hep-th]}}].

\bibitem{El-Showk:2014dwa}
S.~El-Showk, M.~F. Paulos, D.~Poland, S.~Rychkov, D.~Simmons-Duffin and
  A.~Vichi, {{Solving the 3d Ising Model with the Conformal Bootstrap II.
  c-Minimization and Precise Critical Exponents}},
  \href{http://dx.doi.org/10.1007/s10955-014-1042-7}{J. Stat. Phys. {\bf 157},
  869, 2014}, [\href{http://arxiv.org/abs/arXiv:1403.4545}{{arXiv:1403.4545
  [hep-th]}}].

\bibitem{Poland:2010wg}
D.~Poland and D.~Simmons-Duffin, {{Bounds on 4D Conformal and Superconformal
  Field Theories}}, \href{http://dx.doi.org/10.1007/JHEP05(2011)017}{JHEP {\bf
  05}, 017, 2011},
  [\href{http://arxiv.org/abs/arXiv:1009.2087}{{arXiv:1009.2087 [hep-th]}}].

\bibitem{ElShowk:2012hu}
S.~El-Showk and M.~F. Paulos, {{Bootstrapping Conformal Field Theories with the
  Extremal Functional Method}},
  \href{http://dx.doi.org/10.1103/PhysRevLett.111.241601}{Phys. Rev. Lett. {\bf
  111}, 241601, 2013},
  [\href{http://arxiv.org/abs/arXiv:1211.2810}{{arXiv:1211.2810 [hep-th]}}].

\bibitem{Simmons-Duffin:2016wlq}
D.~Simmons-Duffin, {{The Lightcone Bootstrap and the Spectrum of the 3d Ising
  CFT}}, \href{http://dx.doi.org/10.1007/JHEP03(2017)086}{JHEP {\bf 03}, 086,
  2017}, [\href{http://arxiv.org/abs/arXiv:1612.08471}{{arXiv:1612.08471
  [hep-th]}}].

\bibitem{Kos:2016ysd}
F.~Kos, D.~Poland, D.~Simmons-Duffin and A.~Vichi, {{Precision Islands in the
  Ising and $O(N)$ Models}},
  \href{http://dx.doi.org/10.1007/JHEP08(2016)036}{JHEP {\bf 08}, 036, 2016},
  [\href{http://arxiv.org/abs/arXiv:1603.04436}{{arXiv:1603.04436 [hep-th]}}].

\bibitem{Chester:2016wrc}
S.~M. Chester and S.~S. Pufu, {{Towards bootstrapping QED$_{3}$}},
  \href{http://dx.doi.org/10.1007/JHEP08(2016)019}{JHEP {\bf 08}, 019, 2016},
  [\href{http://arxiv.org/abs/arXiv:1601.03476}{{arXiv:1601.03476 [hep-th]}}].

\bibitem{Iliesiu:2018fao}
L.~Iliesiu, M.~Kolo?lu, R.~Mahajan, E.~Perlmutter and D.~Simmons-Duffin, {{The
  Conformal Bootstrap at Finite Temperature}},
  \href{http://dx.doi.org/10.1007/JHEP10(2018)070}{JHEP {\bf 10}, 070, 2018},
  [\href{http://arxiv.org/abs/arXiv:1802.10266}{{arXiv:1802.10266 [hep-th]}}].

\bibitem{Delacretaz:2018cfk}
L.~V. Delacretaz, T.~Hartman, S.~A. Hartnoll and A.~Lewkowycz,
  {{Thermalization, Viscosity and the Averaged Null Energy Condition}},
  \href{http://dx.doi.org/10.1007/JHEP10(2018)028}{JHEP {\bf 10}, 028, 2018},
  [\href{http://arxiv.org/abs/arXiv:1805.04194}{{arXiv:1805.04194 [hep-th]}}].

\bibitem{El-Showk:2016mxr}
S.~El-Showk and M.~F. Paulos, {{Extremal bootstrapping: go with the flow}},
  \href{http://dx.doi.org/10.1007/JHEP03(2018)148}{JHEP {\bf 03}, 148, 2018},
  [\href{http://arxiv.org/abs/arXiv:1605.08087}{{arXiv:1605.08087 [hep-th]}}].

\bibitem{Hellerman:2009bu}
S.~Hellerman, {{A Universal Inequality for CFT and Quantum Gravity}},
  \href{http://dx.doi.org/10.1007/JHEP08(2011)130}{JHEP {\bf 08}, 130, 2011},
  [\href{http://arxiv.org/abs/arXiv:0902.2790}{{arXiv:0902.2790 [hep-th]}}].

\bibitem{Banados:1992wn}
M.~Banados, C.~Teitelboim and J.~Zanelli, {{The Black hole in three-dimensional
  space-time}}, \href{http://dx.doi.org/10.1103/PhysRevLett.69.1849}{Phys. Rev.
  Lett. {\bf 69}, 1849--1851, 1992},
  [\href{http://arxiv.org/abs/arXiv:hep-th/9204099}{{arXiv:hep-th/9204099
  [hep-th]}}].

\bibitem{Friedan:2013cba}
D.~Friedan and C.~A. Keller, {{Constraints on 2d CFT partition functions}},
  \href{http://dx.doi.org/10.1007/JHEP10(2013)180}{JHEP {\bf 10}, 180, 2013},
  [\href{http://arxiv.org/abs/arXiv:1307.6562}{{arXiv:1307.6562 [hep-th]}}].

\bibitem{Collier:2016cls}
S.~Collier, Y.-H. Lin and X.~Yin, {{Modular Bootstrap Revisited}},
  \href{http://dx.doi.org/10.1007/JHEP09(2018)061}{JHEP {\bf 09}, 061, 2018},
  [\href{http://arxiv.org/abs/arXiv:1608.06241}{{arXiv:1608.06241 [hep-th]}}].

\bibitem{Komargodski:2012ek}
Z.~Komargodski and A.~Zhiboedov, {{Convexity and Liberation at Large Spin}},
  \href{http://dx.doi.org/10.1007/JHEP11(2013)140}{JHEP {\bf 11}, 140, 2013},
  [\href{http://arxiv.org/abs/arXiv:1212.4103}{{arXiv:1212.4103 [hep-th]}}].

\bibitem{Fitzpatrick:2012yx}
A.~L. Fitzpatrick, J.~Kaplan, D.~Poland and D.~Simmons-Duffin, {{The Analytic
  Bootstrap and AdS Superhorizon Locality}},
  \href{http://dx.doi.org/10.1007/JHEP12(2013)004}{JHEP {\bf 12}, 004, 2013},
  [\href{http://arxiv.org/abs/arXiv:1212.3616}{{arXiv:1212.3616 [hep-th]}}].

\bibitem{Heemskerk:2009pn}
I.~Heemskerk, J.~Penedones, J.~Polchinski and J.~Sully, {{Holography from
  Conformal Field Theory}},
  \href{http://dx.doi.org/10.1088/1126-6708/2009/10/079}{JHEP {\bf 10}, 079,
  2009}, [\href{http://arxiv.org/abs/arXiv:0907.0151}{{arXiv:0907.0151
  [hep-th]}}].

\bibitem{Mukhametzhanov:2018zja}
B.~Mukhametzhanov and A.~Zhiboedov, {{Analytic Euclidean Bootstrap}},  2018,
  [\href{http://arxiv.org/abs/arXiv:1808.03212}{{arXiv:1808.03212 [hep-th]}}].

\bibitem{Gliozzi:2013ysa}
F.~Gliozzi, {{More constraining conformal bootstrap}},
  \href{http://dx.doi.org/10.1103/PhysRevLett.111.161602}{Phys. Rev. Lett. {\bf
  111}, 161602, 2013},
  [\href{http://arxiv.org/abs/arXiv:1307.3111}{{arXiv:1307.3111 [hep-th]}}].

\bibitem{Gliozzi:2014jsa}
F.~Gliozzi and A.~Rago, {{Critical exponents of the 3d Ising and related models
  from Conformal Bootstrap}},
  \href{http://dx.doi.org/10.1007/JHEP10(2014)042}{JHEP {\bf 10}, 042, 2014},
  [\href{http://arxiv.org/abs/arXiv:1403.6003}{{arXiv:1403.6003 [hep-th]}}].

\bibitem{Li:2017ukc}
W.~Li, {{New method for the conformal bootstrap with OPE truncations}},  2017,
  [\href{http://arxiv.org/abs/arXiv:1711.09075}{{arXiv:1711.09075 [hep-th]}}].

\bibitem{Simmons-Duffin:2015qma}
D.~Simmons-Duffin, {{A Semidefinite Program Solver for the Conformal
  Bootstrap}}, \href{http://dx.doi.org/10.1007/JHEP06(2015)174}{JHEP {\bf 06},
  174, 2015}, [\href{http://arxiv.org/abs/arXiv:1502.02033}{{arXiv:1502.02033
  [hep-th]}}].

\bibitem{Witten:2007kt}
E.~Witten, {{Three-Dimensional Gravity Revisited}},  2007,
  [\href{http://arxiv.org/abs/arXiv:0706.3359}{{arXiv:0706.3359 [hep-th]}}].

\bibitem{Maloney:2007ud}
A.~Maloney and E.~Witten, {{Quantum Gravity Partition Functions in Three
  Dimensions}}, \href{http://dx.doi.org/10.1007/JHEP02(2010)029}{JHEP {\bf 02},
  029, 2010}, [\href{http://arxiv.org/abs/arXiv:0712.0155}{{arXiv:0712.0155
  [hep-th]}}].

\bibitem{Keller:2014xba}
C.~A. Keller and A.~Maloney, {{Poincare Series, 3D Gravity and CFT
  Spectroscopy}}, \href{http://dx.doi.org/10.1007/JHEP02(2015)080}{JHEP {\bf
  02}, 080, 2015},
  [\href{http://arxiv.org/abs/arXiv:1407.6008}{{arXiv:1407.6008 [hep-th]}}].

\bibitem{Hartman:2014oaa}
T.~Hartman, C.~A. Keller and B.~Stoica, {{Universal Spectrum of 2d Conformal
  Field Theory in the Large c Limit}},
  \href{http://dx.doi.org/10.1007/JHEP09(2014)118}{JHEP {\bf 09}, 118, 2014},
  [\href{http://arxiv.org/abs/arXiv:1405.5137}{{arXiv:1405.5137 [hep-th]}}].

\bibitem{Qualls:2013eha}
J.~D. Qualls and A.~D. Shapere, {{Bounds on Operator Dimensions in 2D Conformal
  Field Theories}}, \href{http://dx.doi.org/10.1007/JHEP05(2014)091}{JHEP {\bf
  05}, 091, 2014},
  [\href{http://arxiv.org/abs/arXiv:1312.0038}{{arXiv:1312.0038 [hep-th]}}].

\bibitem{Hellerman:2010qd}
S.~Hellerman and C.~Schmidt-Colinet, {{Bounds for State Degeneracies in 2D
  Conformal Field Theory}},
  \href{http://dx.doi.org/10.1007/JHEP08(2011)127}{JHEP {\bf 08}, 127, 2011},
  [\href{http://arxiv.org/abs/arXiv:1007.0756}{{arXiv:1007.0756 [hep-th]}}].

\bibitem{Keller:2012mr}
C.~A. Keller and H.~Ooguri, {{Modular Constraints on Calabi-Yau
  Compactifications}},
  \href{http://dx.doi.org/10.1007/s00220-013-1797-8}{Commun. Math. Phys. {\bf
  324}, 107--127, 2013},
  [\href{http://arxiv.org/abs/arXiv:1209.4649}{{arXiv:1209.4649 [hep-th]}}].

\bibitem{Chang:2015qfa}
C.-M. Chang and Y.-H. Lin, {{Bootstrapping 2D CFTs in the Semiclassical
  Limit}}, \href{http://dx.doi.org/10.1007/JHEP08(2016)056}{JHEP {\bf 08}, 056,
  2016}, [\href{http://arxiv.org/abs/arXiv:1510.02464}{{arXiv:1510.02464
  [hep-th]}}].

\bibitem{Shaghoulian:2015kta}
E.~Shaghoulian, {{Modular forms and a generalized Cardy formula in higher
  dimensions}}, \href{http://dx.doi.org/10.1103/PhysRevD.93.126005}{Phys. Rev.
  {\bf D93}, 126005, 2016},
  [\href{http://arxiv.org/abs/arXiv:1508.02728}{{arXiv:1508.02728 [hep-th]}}].

\bibitem{Kim:2015oca}
H.~Kim, P.~Kravchuk and H.~Ooguri, {{Reflections on Conformal Spectra}},
  \href{http://dx.doi.org/10.1007/JHEP04(2016)184}{JHEP {\bf 04}, 184, 2016},
  [\href{http://arxiv.org/abs/arXiv:1510.08772}{{arXiv:1510.08772 [hep-th]}}].

\bibitem{Benjamin:2016fhe}
N.~Benjamin, E.~Dyer, A.~L. Fitzpatrick and S.~Kachru, {{Universal Bounds on
  Charged States in 2d CFT and 3d Gravity}},
  \href{http://dx.doi.org/10.1007/JHEP08(2016)041}{JHEP {\bf 08}, 041, 2016},
  [\href{http://arxiv.org/abs/arXiv:1603.09745}{{arXiv:1603.09745 [hep-th]}}].

\bibitem{Das:2017vej}
D.~Das, S.~Datta and S.~Pal, {{Charged structure constants from modularity}},
  \href{http://dx.doi.org/10.1007/JHEP11(2017)183}{JHEP {\bf 11}, 183, 2017},
  [\href{http://arxiv.org/abs/arXiv:1706.04612}{{arXiv:1706.04612 [hep-th]}}].

\bibitem{Cho:2017fzo}
M.~Cho, S.~Collier and X.~Yin, {{Genus Two Modular Bootstrap}},  2017,
  [\href{http://arxiv.org/abs/arXiv:1705.05865}{{arXiv:1705.05865 [hep-th]}}].

\bibitem{Dyer:2017rul}
E.~Dyer, A.~L. Fitzpatrick and Y.~Xin, {{Constraints on Flavored 2d CFT
  Partition Functions}}, \href{http://dx.doi.org/10.1007/JHEP02(2018)148}{JHEP
  {\bf 02}, 148, 2018},
  [\href{http://arxiv.org/abs/arXiv:1709.01533}{{arXiv:1709.01533 [hep-th]}}].

\bibitem{Apolo:2017xip}
L.~Apolo, {{Bounds on CFTs with $\mathcal{W}_3$ algebras and AdS$_3$ higher
  spin theories}}, \href{http://dx.doi.org/10.1103/PhysRevD.96.086003}{Phys.
  Rev. {\bf D96}, 086003, 2017},
  [\href{http://arxiv.org/abs/arXiv:1705.10402}{{arXiv:1705.10402 [hep-th]}}].

\bibitem{Afkhami-Jeddi:2017idc}
N.~Afkhami-Jeddi, K.~Colville, T.~Hartman, A.~Maloney and E.~Perlmutter,
  {{Constraints on higher spin CFT$_{2}$}},
  \href{http://dx.doi.org/10.1007/JHEP05(2018)092}{JHEP {\bf 05}, 092, 2018},
  [\href{http://arxiv.org/abs/arXiv:1707.07717}{{arXiv:1707.07717 [hep-th]}}].

\bibitem{Bae:2017kcl}
J.-B. Bae, S.~Lee and J.~Song, {{Modular Constraints on Conformal Field
  Theories with Currents}},
  \href{http://dx.doi.org/10.1007/JHEP12(2017)045}{JHEP {\bf 12}, 045, 2017},
  [\href{http://arxiv.org/abs/arXiv:1708.08815}{{arXiv:1708.08815 [hep-th]}}].

\bibitem{Collier:2017shs}
S.~Collier, P.~Kravchuk, Y.-H. Lin and X.~Yin, {{Bootstrapping the Spectral
  Function: On the Uniqueness of Liouville and the Universality of BTZ}},
  \href{http://dx.doi.org/10.1007/JHEP09(2018)150}{JHEP {\bf 09}, 150, 2018},
  [\href{http://arxiv.org/abs/arXiv:1702.00423}{{arXiv:1702.00423 [hep-th]}}].

\bibitem{Bae:2018qym}
J.-B. Bae, S.~Lee and J.~Song, {{Modular Constraints on Superconformal Field
  Theories}}, \href{http://dx.doi.org/10.1007/JHEP01(2019)209}{JHEP {\bf 01},
  209, 2019}, [\href{http://arxiv.org/abs/arXiv:1811.00976}{{arXiv:1811.00976
  [hep-th]}}].

\bibitem{Anous:2018hjh}
T.~Anous, R.~Mahajan and E.~Shaghoulian, {{Parity and the modular bootstrap}},
  \href{http://dx.doi.org/10.21468/SciPostPhys.5.3.022}{SciPost Phys. {\bf 5},
  022, 2018}, [\href{http://arxiv.org/abs/arXiv:1803.04938}{{arXiv:1803.04938
  [hep-th]}}].

\bibitem{upcoming}
T.~Hartman, D.~Mazac and L.~Rastelli to appear.

\bibitem{Echeverri:2016ztu}
A.~Castedo~Echeverri, B.~von Harling and M.~Serone, {{The Effective
  Bootstrap}}, \href{http://dx.doi.org/10.1007/JHEP09(2016)097}{JHEP {\bf 09},
  097, 2016}, [\href{http://arxiv.org/abs/arXiv:1606.02771}{{arXiv:1606.02771
  [hep-th]}}].

\bibitem{Gorbenko:2018ncu}
V.~Gorbenko, S.~Rychkov and B.~Zan, {{Walking, Weak first-order transitions,
  and Complex CFTs}}, \href{http://dx.doi.org/10.1007/JHEP10(2018)108}{JHEP
  {\bf 10}, 108, 2018},
  [\href{http://arxiv.org/abs/arXiv:1807.11512}{{arXiv:1807.11512 [hep-th]}}].

\bibitem{Arkani-Hamed:2018ign}
N.~Arkani-Hamed, Y.-T. Huang and S.-H. Shao, {{On the Positive Geometry of
  Conformal Field Theory}},  2018,
  [\href{http://arxiv.org/abs/arXiv:1812.07739}{{arXiv:1812.07739 [hep-th]}}].

\bibitem{pinkus1996spectral}
A.~Pinkus, {Spectral properties of totally positive kernels and matrices},  in
  \emph{Total positivity and its applications}, pp.~477--511.
\newblock Springer, 1996.

\bibitem{Belin:2014fna}
A.~Belin, C.~A. Keller and A.~Maloney, {{String Universality for Permutation
  Orbifolds}}, \href{http://dx.doi.org/10.1103/PhysRevD.91.106005}{Phys. Rev.
  {\bf D91}, 106005, 2015},
  [\href{http://arxiv.org/abs/arXiv:1412.7159}{{arXiv:1412.7159 [hep-th]}}].

\bibitem{Vafa:2005ui}
C.~Vafa, {{The String landscape and the swampland}},  2005,
  [\href{http://arxiv.org/abs/arXiv:hep-th/0509212}{{arXiv:hep-th/0509212
  [hep-th]}}].

\end{thebibliography}\endgroup
\bibliographystyle{ourbst}

\end{document}